\documentclass[twocolumn]{emulateapj}
\usepackage{graphicx}
\usepackage{dcolumn}

\def\arcsec{\hbox{$^{\prime\prime}$}}

\newcommand       \pc           {\,{\rm pc}}

\newcommand{\vcsigma}{v$_{{c}}$\,-- \,$\sigma$}
\newcommand{\MBH}{M$_{{BH}}$}

\newcommand{\vc}{v_{c}}
\newcommand{\MBHsigma}{M$_{{BH}}$\, --\, $\sigma$}

\newcommand{\kms}{{{km\,s$^{-1}$}}}
\newcommand{\Msun}{{{M$_{\odot}$}}}

\begin{document}

\title{The Discovery of an Active Galactic Nucleus in the Late-type Galaxy NGC 3621: {\it 
Spitzer} Spectroscopic Observations}

\author{S. Satyapal\altaffilmark{1}, D. Vega\altaffilmark{1}, T. Heckman\altaffilmark{2}, 
B. O'Halloran \altaffilmark{1} \& R. Dudik\altaffilmark{1,3}}

\altaffiltext{1}{George Mason University, Department of Physics \& Astronomy, MS 3F3, 4400 
University Drive, Fairfax, VA 22030; satyapal@physics.gmu.edu}

\altaffiltext{2}{Center for Astrophysical Sciences, Department of Physics and Astronomy, The Johns Hopkins University, Baltimore, MD 21218}

\altaffiltext{3}{Observational Cosmology Laboratory, Code 665, NASA Goddard Space Flight Center, Greenbelt, MD 20771}

\begin{abstract}
We report the discovery of an Active Galactic Nucleus (AGN) in the nearby SAd galaxy NGC 
3621 using {\it Spitzer} high spectral resolution observations.  These observations reveal 
the presence of [NeV] 14 $\mu$m and 24 $\mu$m emission which is centrally concentrated and 
peaks at the position of the near-infrared nucleus.  Using the [NeV] line luminosity, we 
estimate that the nuclear bolometric luminosity of the AGN is $\sim$ 5$\times$10$^{41}$ 
ergs s$^{-1}$, which corresponds based on the Eddington limit to a lower mass limit of the 
black hole of $\sim$  4$\times$10$^3$\Msun.  Using an order of magnitude estimate for the 
bulge mass based on the Hubble type of the galaxy, we find that this lower mass limit does 
not put a strain on the well-known relationship between the black hole mass and the host 
galaxy's stellar velocity dispersion established in predominantly early-type galaxies.  
Mutliwavelength follow-up observations of NGC 3621 are required to obtain more precise 
estimates of the bulge mass, black hole mass, accretion rate, and nuclear bolometric 
luminosity.  The discovery reported here adds to the growing evidence that a black hole {\it can} 
form and grow in a galaxy with no or minimal bulge.

\end{abstract}


\section{Introduction}

The discovery that at the heart of virtually all early-type galaxies in the local Universe, 
lie supermassive nuclear black holes (SBHs) has led to the general consensus that black 
holes play a pivotal role in the formation and evolution of galaxies.  The well-known 
correlation between the black hole mass, M$_{\rm BH}$, and the host galaxy's stellar 
velocity dispersion, $\sigma$ (Gebhardt et al. 2000; Ferrarese \&  Merritt 2000) implies 
that black hole growth and the build-up of galaxy bulges go hand-in-hand, perhaps as 
feedback from the active galactic nucleus (AGN) regulates the surrounding star formation in 
the host galaxy (e.g. Silk \& Rees 1998; Kauffmann \& Haehnelt 2000).  The connection 
between black hole growth and galaxy bulges is further intimated by the fact that virtually 
all {\it currently} known actively accreting black holes  i.e. AGN - in the local Universe 
are found in galaxies with prominent bulges (e.g. Heckman 1980a; Ho, Filippenko, \&  Sargent 1997; Kauffmann et al. 2003). It is 
unclear how common it is for bulgeless galaxies to contain SBHs and of those that do, 
whether they are actively accreting.  M33, the best studied nearby bulgeless galaxy, shows 
no evidence of a SBH and the upper limit on the mass is significantly below that predicted 
by the M$_{\rm BH}$-$\sigma$ relation established in early-type galaxies (Gebhardt et al. 
2001; Merritt et al. 2001).

Amongst the population of known AGN, there are possibly a handful of extremely late-type 
galaxies that may show subtle signs of AGN activity in their optical narrow-line nuclear 
spectra (Ho et al. 1997).  The best-studied definitive case of an AGN in a purely bulgeless 
galaxy is the galaxy NGC 4395, which shows an unmistakable Seyfert 1 spectrum. The inferred 
black hole mass in this dwarf galaxy is 3.6 $\times$ 10$^5$ M$_{\odot}$ (Peterson et al. 
2005), much less massive than black holes found in galaxies with massive bulges, and comparable to the inferred black hole mass in the other well-know dwarf galaxy with an AGN, POX 52 (Barth et al. 2004).  Greene \& 
Ho (2004) recently searched the First Data Release of the Sloan Digital Sky Survey for 
galaxies with similar intermediate mass black holes and found only 19 broad-line AGN, 
suggesting that they are uncommon.  Subsequent stellar velocity dispersion measurements 
revealed that these objects follow the extrapolation of the M$_{\rm BH}$-$\sigma$ relation 
(Barth et al. 2005).  It is unclear if the hosts are late-type galaxies since the Sloan 
images are of insufficient spatial resolution to confirm their morphological type.

Are AGN uncommon in bulgeless galaxies?  Is a bulge necessary for a black hole to form and 
grow? These questions cannot be definitively answered with the current suite of 
observations, which are largely carried out at optical wavelengths.  Such studies can be 
severely limited in the study of bulgeless galaxies, where a putative AGN is likely to be both energetically weak and deeply embedded in the center of a dusty late-type spiral.  In such systems, the traditional optical emission lines used to identify AGN can be dominated by emission from star formation regions, in addition to being significantly attenuated by dust in the host galaxy.  Indeed searching for AGN under these circumstances poses unique challenges.  Although X-ray observations can be a powerful tool in finding optically obscured AGN in the non-Compton-thick regime, the X-ray luminosities of weak AGN will be low and comparable to and therefore indistinguishable from X-ray binaries in the host galaxy.  Similarly, the radio emission can be dominated by and indistinguishable from compact nuclear starbursts (e.g., Condon et al., 1991). {\it Spitzer} mid-IR spectroscopy is the ideal tool to search 
for AGN in such galaxies. As has been shown in previous works, AGN show prominent high excitation fine 
structure line emission but starburst and normal galaxies are characterized by a lower 
excitation spectra characteristic of HII regions ionized by young stars (e.g., Genzel et al. 
1998; Sturm et al. 2002; Satyapal et al. 2004).  In particular, the [NeV] 14 $\mu$m  
(ionization potential 96 eV) line is not produced in HII regions surrounding young stars, 
the dominant energy source in starburst galaxies, since even hot massive stars emit very 
few photons with energy sufficient for the production of this ion.  The detection of this 
line in a galaxy is therefore definitive proof of an AGN.

In this {\it Letter}, we report the discovery of an AGN in the galaxy NGC 3621.  NGC 3621 
is a relatively isolated nearby (6.2 Mpc; Rawson et al. 1997) SAd III-IV (Third Reference 
Catalogue of Bright Galaxies  RC3; de Vaucouleurs et al. 1991) galaxy with no previous 
published evidence for nuclear activity.  The observations presented here add to the 
growing evidence that a black hole can form and grow in a galaxy with little or no bulge.

\section{Observations and Data Reduction}

NGC 3621 was observed by the Infrared Spectrometer (Houck et al. 2004) on board {\it 
Spitzer} using the short-wavelength high-resolution (SH; 4.7\arcsec$\times$11.3\arcsec, 9.9-19.6 $\mu$m) and the long-wavelength high-resolution (LH; 11.1\arcsec$\times$22.3\arcsec, 18.7-37.2 $\mu$m) modules as part of the SINGS Legacy 
Proposal (program ID 159; Kennicutt et al. 2003) on 2004 June 28. These modules have a 
spectral resolution of R $\sim$ 600.  The observations were executed in spectral mapping 
mode, in which the spacecraft moves in a raster pattern of discrete steps in order to 
construct a rectangular map of the targeted region.  The SH and LH maps included 3 
pointings parallel to and 5 pointings perpendicular to the major axis of the slit, with 
half slit-length and half-slit width steps, respectively.  The integration time per 
pointing was approximately 60s, with each position being covered twice for the SH 
observations.  The total duration for the high resolution observations of NGC 3621 was 
2915s.

We used BCD-level data products downloaded from the {\it Spitzer} archive in conjunction 
with \emph{CUBISM} v.1.0.2\footnote[3]{URL: 
http://ssc.spitzer.caltech.edu/archanaly/con\-tri\-bu\-ted/cu\-bism/in\-dex.html} (Kennicutt 
et al. 2003; Smith et al. 2004) to construct the high-resolution spectral cubes for 
NGC~3621. The BCD-level products were pre-processed by the {\it Spitzer} pipeline, version 
13.2\footnote[4]{{\it Spitzer} Observers Manual, URL: 
http://ssc.spitzer.cal\-tech.edu/doc\-uments/som/} prior to download.  The overall flux calibration uncertainty is 25 to 30\%.  A detailed 
description of the post-processing steps included in {\it CUBISM} is given in Smith et al. 
(2004).

The final full cube map size for SH corresponds to 
$\sim$24.7\arcsec$\times$13.5\arcsec, while the final full cube map size for LH corresponds 
to $\sim$44.6\arcsec$\times$30.8\arcsec.  At the distance of NGC 3621, this corresponds to a physical size of 742pc$\times$405pc and 1341\pc$\times$923\pc.  Given the small spatial extent of 
both the SH and LH maps, we were unable to perform in-situ background subtraction, since the 
full extent of each map is confined within the galaxy.  

The post-BCD software SMART, v.5.5.7 (Higdon et al., 2004), was then used to 
obtain line fluxes from the extracted 1D spectra. When line ratios were calculated, the line flux was 
obtained from spectra extracted from the same-sized aperture 
in both modules corresponding to the same physical region 
on the galaxy.  This aperture can range from a minimum of $\sim$ 9\arcsec$\times$9\arcsec, below which artifacts introduced by an undersampled PSF can distort the spectrum, to a maximum of 22.6\arcsec$\times$14.8\arcsec\ -- i.e., roughly the full extent of the SH map.  The FITS output images from \emph{CUBISM} 
were smoothed within \emph{ds9} using a Gaussian kernel of 2-pixel width.  Finally, the line map plots were created using the IRAF routine 
\emph{rotate} in conjunction with \emph{ds9}.

\section{Results and Discussion}

In Figure 1, we show spectra extracted from the full aperture of the SH and LH maps near 14 
$\mu$m,  24 $\mu$m, and 26$\mu$m.   As can be seen, there are clear detections of the 
[NeV]14$\mu$m (4$\sigma$), [NeV]24$\mu$m (6$\sigma$), and the [OIV] 25.9$\mu$m (10$\sigma$) 
lines, providing strong evidence for the presence of an AGN. Continuum-subtracted spectral 
images reveal that the emission is centrally concentrated and peaks at the position of the 
nucleus determined from the 2MASS coordinates, as can be seen in Figure 2.   Since the 
ratio of high to low excitation lines depends on the nature of the ionizing source, the 
[NeV]14$\mu$m/[NeII]12.8$\mu$m and the [OIV]25.9$\mu$m/[NeII]12.8$\mu$m line flux ratios 
have been used to characterize the nature of the dominant ionizing source in galaxies 
(e.g., Genzel et al. 1998; Sturm et al. 2002; Satyapal et al. 2004; Dale et al. 2006).  The 
[NeV]/[NeII] ratio for the 13 AGN with both [NeV] and [NeII] detections by the {\it 
Infrared Space Observatory (ISO)} ranges from 0.06 to 2.11, with a median value of 0.47 
(Sturm et al. 2002).  The [NeV]/[NeII] ratio corresponding to the maximum aperture (22.6\arcsec$\times$14.8\arcsec\ )for this galaxy is 0.06, similar to the lowest 
value observed in a similar aperture (27\arcsec$\times$14\arcsec\ ) by {\it ISO}. The [NeV]/[NeII] ratio  does not increase substantially as the aperture size is reduced; the line ratio corresponding to the minimum aperture (9\arcsec$\times$9\arcsec\ ) is 0.064.  The [OIV]/[NeII] ratio for the 17 AGN with both lines 
detected by {\it ISO} ranges from 0.15 to 8.33, with a median value of 1.73 (Sturm et al. 
2002).  The [OIV]/[NeII] line flux ratio corresponding to the maximum aperture in NGC 3621 is 0.23, again within the range  but at the low end of the observed values in the nearby powerful AGN observed by {\it ISO}. The [OIV]/[NeII] flux ratio corresponding to the minimum aperture (9\arcsec$\times$9\arcsec) is 0.3, only marginally larger than the ratio obtained from the larger aperture.  For comparison, the few 
starburst galaxies that show detectable [OIV] emission have [OIV]/[NeII] line flux ratios 
that range from 0.006 to 0.647 (median = 0.019; Verma et al. 2003) but no [NeV] emission.  
We note that the fact that the [NeV]/[NeII] and [OIV]/[NeII] line flux ratios are very low 
suggests that the {\it Spitzer} spectrum is dominated by regions of star formation, and that significant contamination from star formation exists even at the smallest {\it Spitzer} apertures (9\arcsec$\times$9\arcsec $\sim$ 270\pc$\times$270\pc)  We note that there is currently no published optical spectrum of NGC 3621.  However, an optical spectrum has been obtained by the SINGS team which does show a very low-power Seyfert, but it becomes inconspicuous for an aperture size of $\sim$ 1kpc due to contamination from star formation in the host galaxy (Moustakas et al. 2007 in prep.).  In the general case, for such low luminosity AGN, the standard optical or UV emission lines can be ambiguous indicators of AGN 
activity because they can all be produced in starburst models (e.g., Terlevich et al. 
1992), a problem that is exacerbated in weak AGN.  However, the detection of the two [NeV] 
lines by {\it Spitzer} provides firm evidence for an AGN in this galaxy and demonstrates 
that {\it Spitzer} {\it can} find AGN in galaxies even when the AGN is energetically minor 
compared to star formation, regardless of the aperture size from which the spectrum is obtained.

\begin{figure*}[htbp]
\centering
\begin{tabular}{ccc}
  \includegraphics[width=0.30\textwidth]{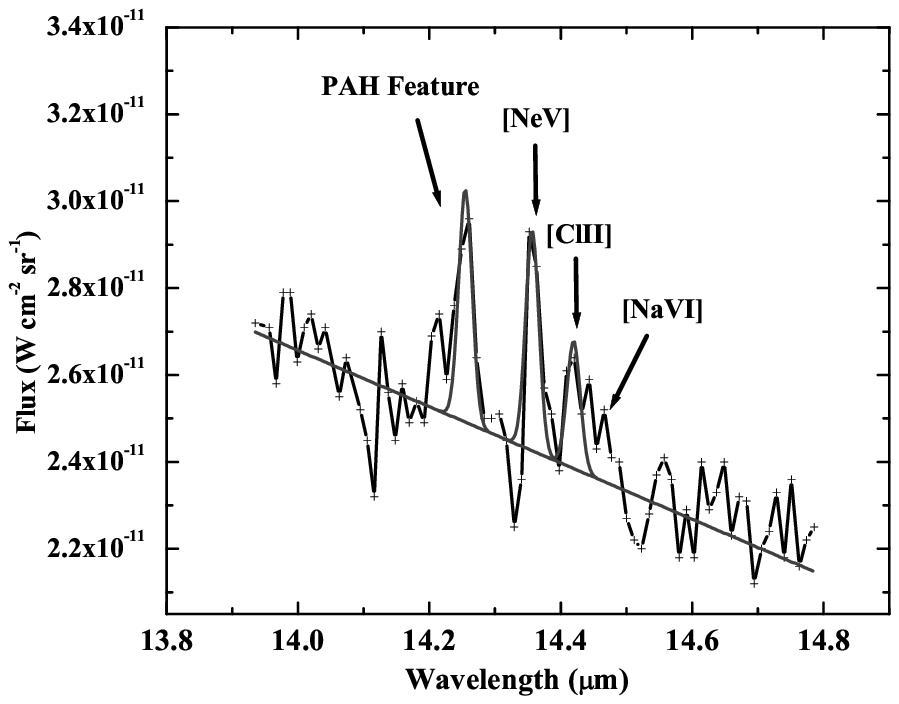} &
  \includegraphics[width=0.30\textwidth]{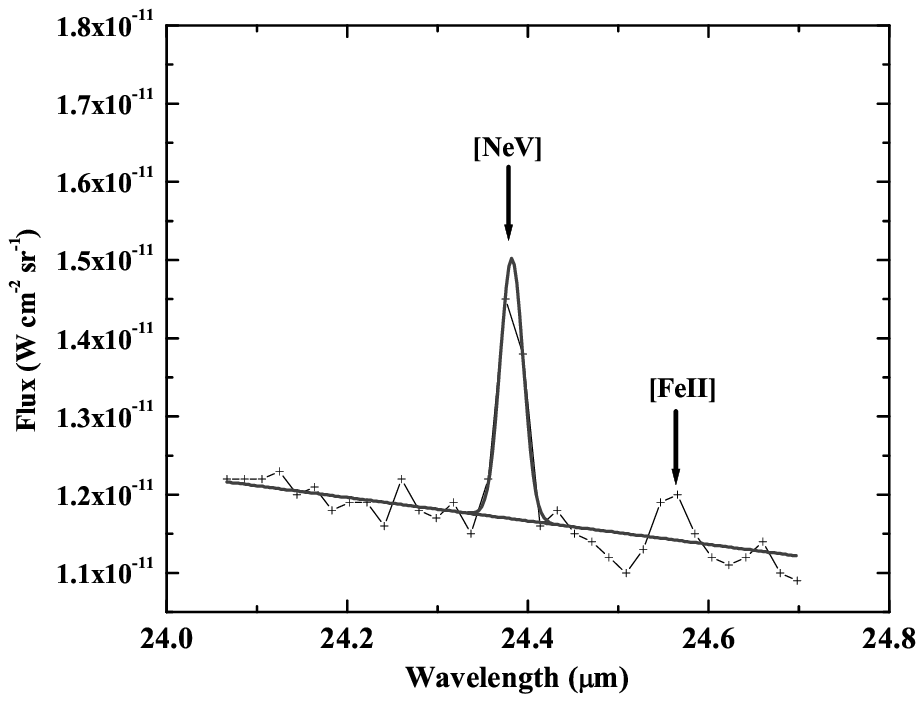} &
  \includegraphics[width=0.30\textwidth]{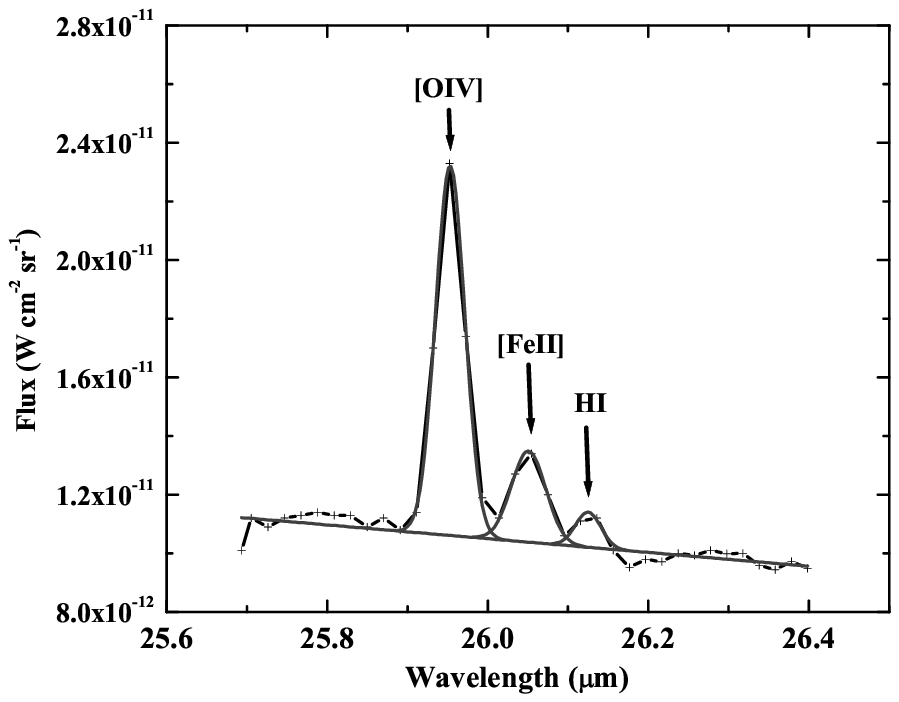} \\
  \includegraphics[width=0.30\textwidth]{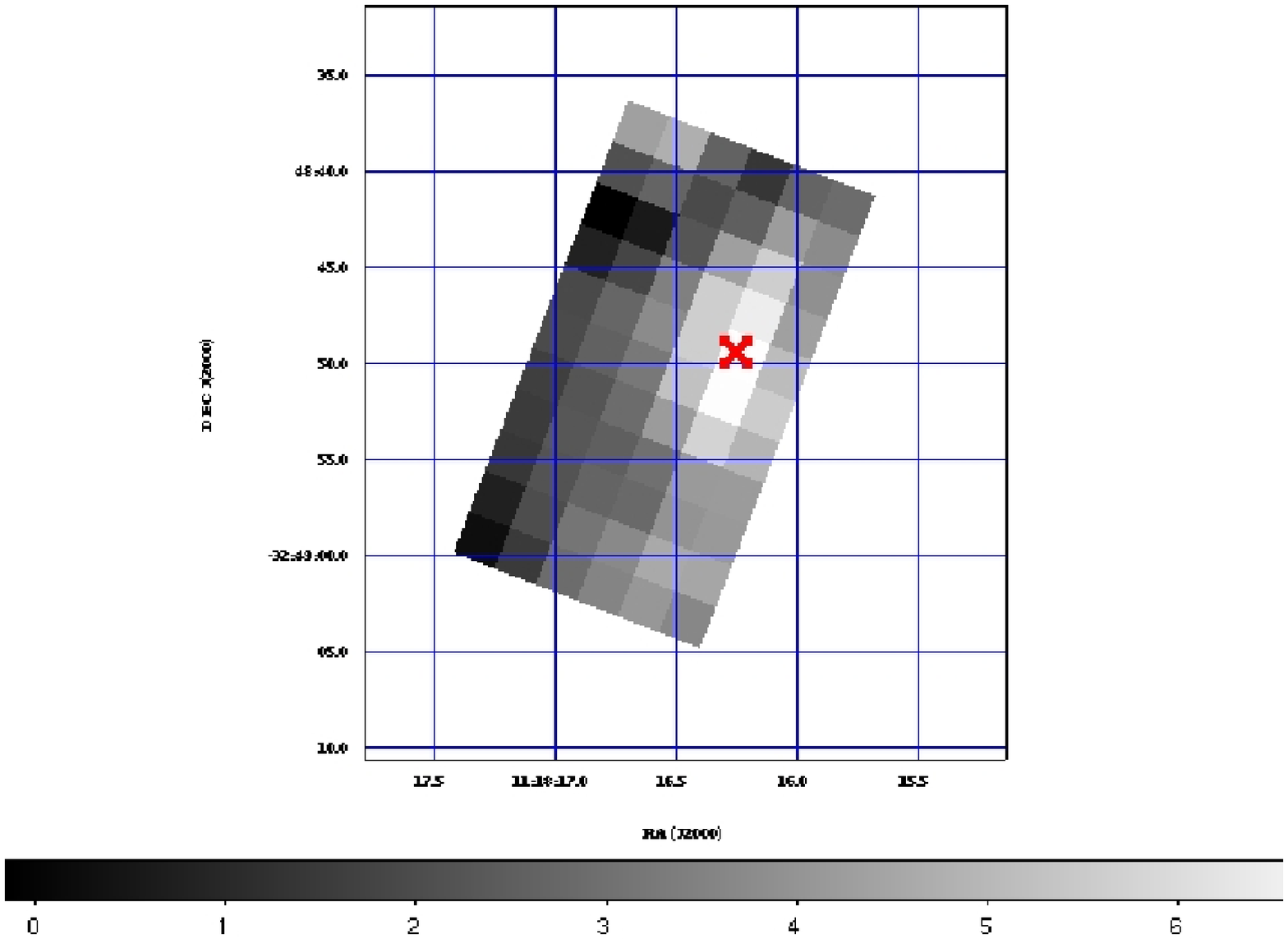} &
  \includegraphics[width=0.30\textwidth]{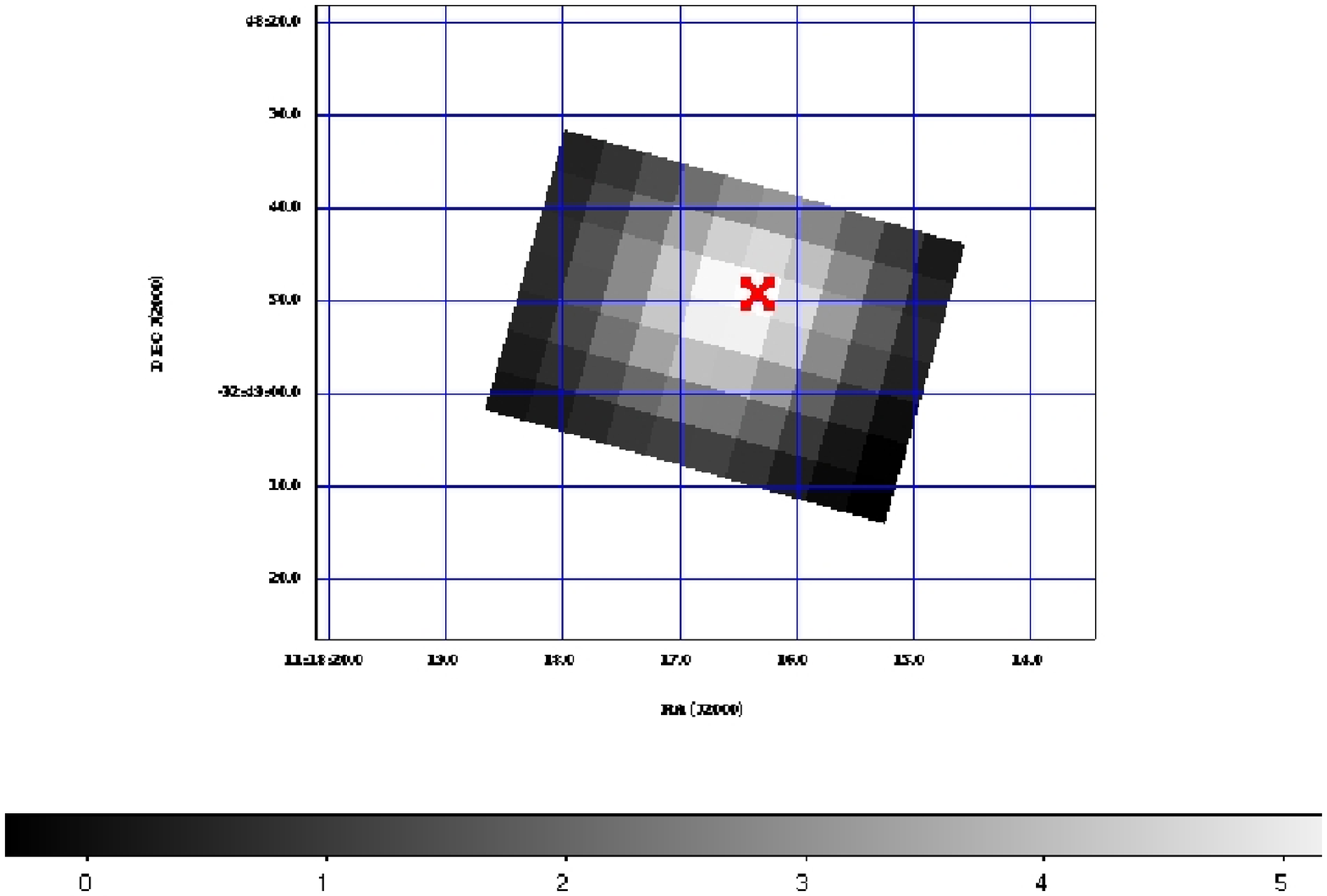} &
  \includegraphics[width=0.30\textwidth]{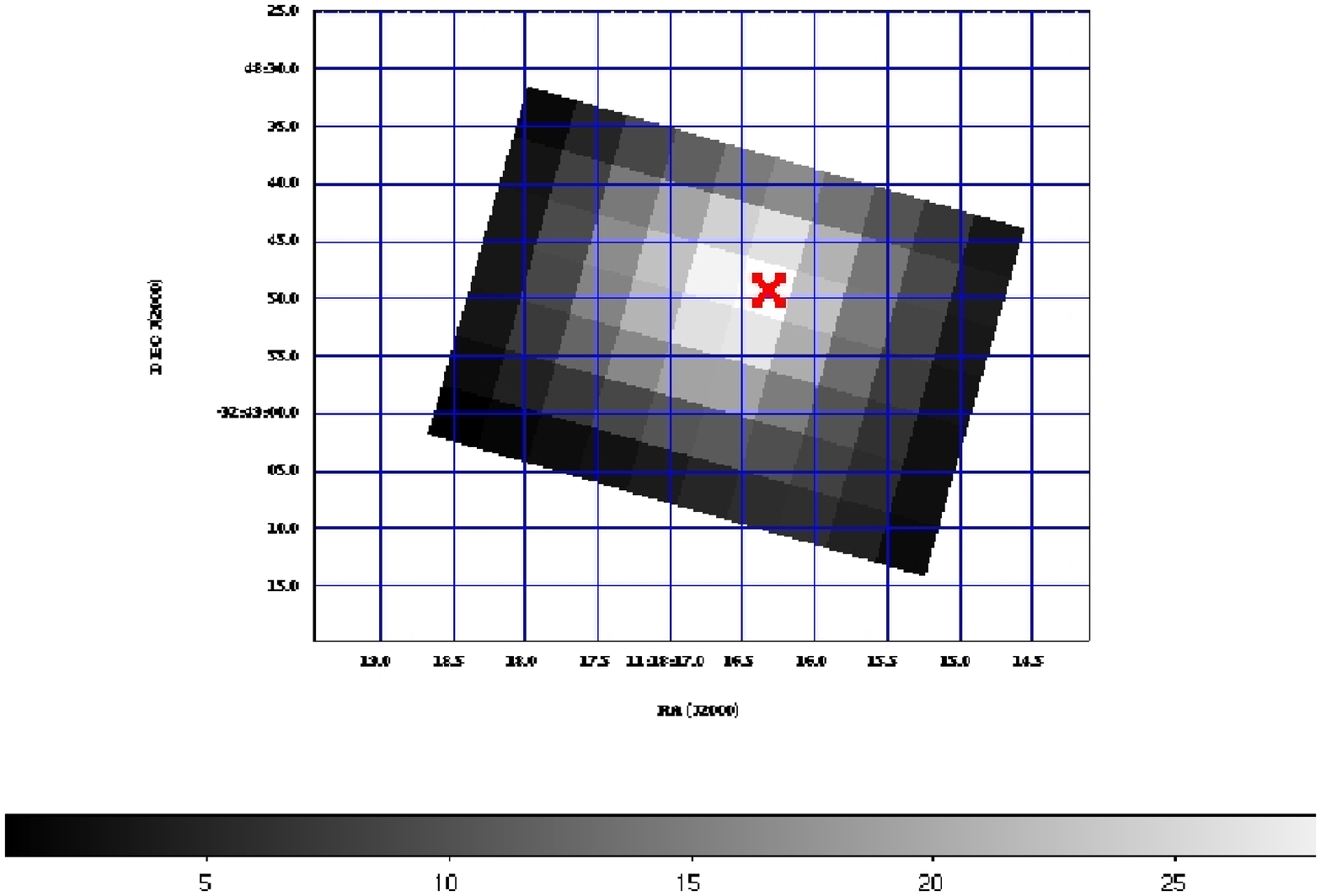} \\
\end{tabular}
\caption[]{(a) SH spectrum showing the detection of the [NeV] 14$\mu$m line. The flux of 
the line is (8.66$\pm$4.06)$\times$10$^{-22}$ W cm$^{-2}$.(b) LH spectrum showing the 
detection of the [NeV] 24$\mu$m line. The flux of the line is 
(8.56$\pm$2.88)$\times$10$^{-22}$ W cm$^{-2}$. (c) LH spectrum showing the detection of the 
[OIV] 26$\mu$m line. The flux of the line is (4.12$\pm$0.87)$\times$10$^{-21}$ W cm$^{-2}$. 
Continuum-subtracted images of the (d)[NeV] 14$\mu$m line, (e) [NeV] 24$\mu$m line, and 
(f)[OIV] 26$\mu$m line. All spectra shown in the figure are extracted from  a 23\arcsec 
$\times$ 15\arcsec aperture centered on the nucleus. The cross indicates the nuclear 
coordinates from the 2MASS database. Note that in all images the emission is concentrated 
and centered on the nucleus.
}
\end{figure*}

\subsection{ Estimating the Bolometric Luminosity of the AGN and Limits on the Black Hole 
Mass}

Since there are currently no published optical spectroscopic observations of NGC 3621, it is 
not possible to estimate the bolometric luminosity of the AGN using traditional 
optical calibration factors (e.g. Kaspi et al. 2000).  However, we can get an order of 
magnitude estimate of the bolometric luminosity of the AGN using the [NeV] line luminosity 
since this line is not contaminated by emission from the host galaxy and can therefore be 
assumed to be associated exclusively with the AGN.  Using the [NeV] 14$\mu$m fluxes from a 
large sample of AGN recently observed by {\it Spitzer} (Dudik et al. 2007; Gorjian et al. 
2007), we can determine the relationship between the line luminosity and the nuclear 
bolometric luminosity of the AGN.  Selecting only those galaxies with published bolometric 
luminosities obtained through direct integration of a well-sampled nuclear spectral energy 
distribution (SED), we plot in Figure 2, L$_{\rm [NeV]}$ vs. L$_{\rm bol}$, demonstrating 
that there is a clear correlation.  The bolometric luminosities for this sample ranged from $\sim$ 2$\times$10$^{43}$ ergs s$^{-1}$ to 4$\times$10$^{46}$ ergs s$^{-1}$ and the black hole masses ranged from $\sim$  7$\times$10$^6$\Msun to 7$\times$10$^9$\Msun. The best-fit linear relation yields:
\begin{equation}
\log(L_{\rm bol}) =  (0.938)\log(L_{\rm [NeV]}) + 6.317
\end{equation}

Assuming this relationship extends to lower values of L$_{\rm [NeV]}$, the total [NeV] 
14$\mu$m luminosity from the entire map of NGC 3621 of $\sim$ 5$\times$10$^{37}$ ergs 
s$^{-1}$ corresponds to a nuclear bolometric luminosity of $\sim$ 5$\times$10$^{41}$ ergs 
s$^{-1}$.  If we assume that the AGN is radiating at the Eddington limit, this yields a 
lower limit to the mass of the black hole of $\sim$  4$\times$10$^3$\Msun.

\begin{figure}[]
\begin{center}
\noindent{\includegraphics[width=7cm]{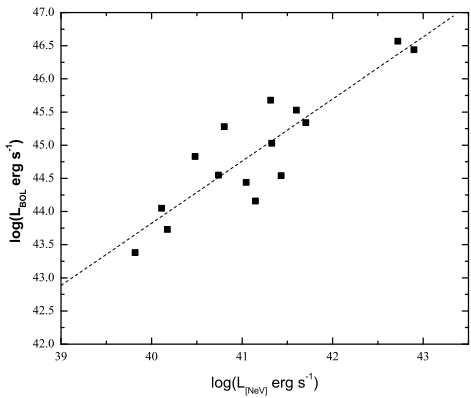}}
\end{center}
\caption[]{The [NeV] line luminosity as a function of nuclear bolometric luminosity in 
known AGN that currently have [NeV] observations (Dudik et al. 2007; Gorjian et al. 2007).  
The bolometric luminosities were taken from Woo \& Urry (2002) and were obtained by direct 
flux integration of a well-sampled SED.  The figure clearly shows a strong correlation, 
demonstrating that the [NeV] luminosity can be used to estimate the bolometric luminosity 
of the AGN in the proposed sample.
}
\end{figure}

Does this lower mass limit allow us to make any statement on the location of NGC 3621 on 
the \MBHsigma~plane?   There are no previously published high spectral resolution optical 
observations from which the stellar velocity dispersion in this galaxy can be determined.  
In the absence of an explicit bulge-disk decomposition from the surface photometry, we can 
get an order of magnitude estimate of the bulge luminosity using the morphological type of 
the galaxy and its total luminosity.  We adopt the empirical relation given in Simien \& de 
 Vaucouleurs (1986) to estimate the contribution of the disk to the total luminosity of a 
disk galaxy: If --3 $\leq$ T $\leq$ 7: $\Delta m_{\rm bul}$ = 0.324(T + 5) -- 0.054(T + 
5)$^2$ + 0.0047(T + 5)$^3$, where T is the numerical Hubble type index, as given in the 
RC3.  The bulge absolute magnitudes is then given by: $M_B$(bul) = $M_{B_T}^0$ + $\Delta 
m_{\rm bul}$.

Using the apparent magnitude from the RC3 catalog of B$_{T}^{0}$=9.20 and Hubble index of 7, yields 
$M_B$(bul)=--15.52 mag.  Using the updated calibration of the Magorrian relationship from 
Ferrarese \& Ford (2005), the expected black hole mass is \MBH = 3.1$\times$10$^6$ \Msun, 
approximately three orders of magnitude higher than the lower mass limit derived from the 
Eddington limit.  Since the scatter in the \MBH - M$_{{B}}$ relation is significantly 
higher than that in the \MBHsigma   relation (e.g. Ferrarese \& Merritt 2000), we can attempt 
to get an estimate of $\sigma$ using the maximum rotational velocity, $\vc$.  The maximum 
rotational velocity has been shown to follow a tight correlation with the stellar velocity 
dispersion in spiral galaxies in which extended rotation curves are available (Ferrarese 
2002; Baes et al. 2003; Courteau et al. 2007).  As part of the SINGS data deliveries, H$\alpha$ rotation curves 
are available for NGC 3621 (Daigle et al. 2006) \footnote[5]{rotation curves are publically 
available at URL: http://www.astro.umontreal.ca/fantomm/sings/rotation$_{}$curves.htm}.  The 
maximum rotational velocity obtained from this rotation curve is 180 \kms, which assuming 
an inclination angle of 65 degrees (Daigle et a. 2006) corresponds to an inclination-corrected velocity of $\sim$ 200 \kms.  However, the H$\alpha$ data cover only the inner 
$\sim$6 kpc of the galaxy, which unfortunately does not extend to the flat part of the 
rotation curve.  Noting that this is an underestimate of $\vc$, the value of $\sigma$ obtained using the \vcsigma~relation from Courteau et al. 2007, shown in spiral galaxies to depend on the bulge-to-total (B/T) light ratio, is $\sim$ 102 km/s .  The corresponding black hole mass is $\sim$ 6$\times$10$^6$ \Msun, again 
significantly larger than the lower mass limit derived from the Eddington limit.  We note that all of these methods to estimate the black hole mass are indirect and subject to large uncertainties.  Some spiral galaxies of Hubble type Sd can contain nuclear star clusters and no bulge, some have bulges and no nuclear star clusters, and some possess both (Boker, Stanek, \& van der Marel 2003).  A detailed high--spatial resolution surface brightness profile analysis is required to determine the bulge and nuclear star cluster content in NGC 3621.  

\section{Implications}

The discovery of an AGN in NGC 3621 adds to the growing evidence that a black hole can form 
and grow in a galaxy with no or minimal bulge component.  Our estimate for the bolometric 
luminosity of the AGN is approximately two orders of magnitude greater than that of NGC 
4395 (Peterson et al. 2005), the best-studied example of a bulgeless galaxy hosting an AGN. 
 These observations suggests that perhaps it is not the bulge but the dark matter halo of 
the host galaxy that determines the presence and activity level of SBHs, as suggested by 
recent theoretical models for the formation of supermassive black holes (e.g., Adams, 
Graff, \& Richstone 2001; Natarajan, Sigurdsson, \& Silk 1998).  Follow-up multiwavlength observations of 
NGC 3621 are required to obtain more precise estimates of the bulge mass, black hole mass, 
accretion rate, and nuclear bolometric luminosity.

Are AGN in bulgeless galaxies more common than once thought?  Since the optical emission lines in weak AGN may be dominated by star formation in the host galaxy, it is possible that AGN are missed in optical spectroscopic surveys and that they are perhaps more common than 
current optical surveys suggest.   Moreover, it is well known that the infrared (IR) excess observed 
in galaxies increases along the Hubble Sequence implying that late-type galaxies are dusty 
(e.g. Sauvage \& Thuan 1994).  In order to truly 
determine how common SBHs and AGN activity are in bulgeless galaxies and whether or not the 
dark matter mass instead of the bulge mass determines the presence and activity level of a 
SBH, an infrared spectroscopic study is crucial.  Future studies with {\it Spitzer} will 
allow us to gain further insight into these fundamental questions and expand our 
understanding of the formation and growth of SBHs in low bulge environments.
\acknowledgements
We are very thankful to Joel Green , Dan Watson, Jackie Fischer, Mario Gliozzi, Mike Eracleous, and Rita Sambruna for their invaluable help in the data 
analysis, and for their enlightening and thoughtful comments.  The excellent suggestions of an anonymous referee helped improve this Letter. SS gratefully acknowledges financial 
support from NASA grant NAG5-11432.  RPD gratefully acknowledges financial support from the NASA Graduate Student Research Program.

\end{document}